\documentclass[10pt]{article}

\usepackage{fullpage, a4wide}
\usepackage[T1]{fontenc}
\usepackage[utf8]{inputenc}
\usepackage{amssymb}
\usepackage{amsmath}
\usepackage{graphicx}
\usepackage{enumitem}
\usepackage{cite}

\newcommand{\difftm}{Diff-Thue-Morse }

\newtheorem{theorem}{Theorem}

\newtheorem{lemma}[theorem]{Lemma}

\title{A Separation Between Run-Length SLPs and LZ77}

\author{Philip Bille \and Travis Gagie \and Inge Li G{\o}rtz \and Nicola Prezza}
\begin{document}
\maketitle

\begin{abstract}
In this paper we give an infinite family of strings for which the length of the Lempel-Ziv'77 parse is a factor $\Omega(\log n/\log\log n)$ smaller  than the smallest run-length grammar.
\end{abstract}

\section{Introduction}

The Lempel-Ziv factorization of a text~\cite{lempel1976complexity} (LZ77) is a greedy left-to-right parse in maximal factors such that each factor already occurred to the left. Despite its simplicity, LZ77 can be easily shown to be optimal among all unidirectional parses (i.e. that copy phrases from left-to-right), and dominates other popular compression schemes such as Straight Line Programs (SLPs), i.e., context-free grammars that generate only the text as output.
Let $z_{no}$ be the number of phrases of the Lempel-Ziv parse when overlaps are not allowed between phrases and their sources, and let $g^*$ be the size of the smallest SLP.  Charikar at al.~\cite{charikar2005smallest} and Rytter~\cite{rytter2003application} showed how to obtain a unidirectional parse of size at most $g$ starting from a SLP of size $g$. It follows from the optimality of LZ77 that the relation $z_{no}\leq g^*$ holds. On the other hand, Charikar at al.~\cite{charikar2005smallest} showed an infinite family of strings for which $g^*/z_{no} = \Omega(\log n/\log\log n)$, where $n$ is the length of the string. Together, these results imply that LZ77 compression without overlaps is always at least as good as grammar compression, and strictly better in some cases. 

Given that, in fields such as compressed computation, SLPs are often easier to treat than LZ77, one might wonder whether we could enhance SLPs so that they become as powerful as Lempel-Ziv compression. See, for example, Bille et al.~\cite[Thm 1.1]{bille2015random} and Kreft ad Navarro~\cite[Thm 4.11]{kreft2013compressing} for classical solutions to the random access problem on grammar- and Lempel-Ziv-compressed texts, respectively. One possible extension of SLPs is to add so-called run-length rules, i.e. rules of the form $X\rightarrow Y^\ell$, for $\ell>1$ (meaning that $X$ expands to $\ell$ repetitions of $Y$). This extension takes the name \emph{run-length SLP}, or RLSLP in what follows~\cite{nishimoto2016fully}. Let $g^*_{rl}$ be the size of the smallest RLSLP. 
It is easy to show that $g^* = \Theta(\log n)$ and $g^*_{rl} = O(1)$ on unary strings of length $n$. This implies that RLSLPs are a strict improvement over SLPs. Since $z_{no} \in \Theta(\log n)$ on unary strings, we also have that $z_{no}/g^*_{rl} = \Theta(\log n)$ for an infinite class of strings: RLSLPs improve upon Lempel-Ziv compression in some cases, and therefore are good candidates for capturing it. 
However, a slight modification to the LZ77 compression scheme adds enough power to capture, again, grammar compression with run-length rules. Let $z$ be the number of phrases of the Lempel-Ziv parse when overlaps are allowed between phrases and their sources. By adapting Rytter's proof, Gagie et al. in~\cite{gagie2017optimal} proved that $z \leq g^*_{rl}$, which implies that
we cannot hope to beat LZ77 with overlaps using RLSLPs. 

The missing piece in the puzzle is the following: are RLSLPs always at least as good as Lempel-Ziv (with or without overlaps)? In this paper, we answer negatively to this question. By adapting Charikar at al.'s proof~\cite{charikar2005smallest}, we give an infinite family of strings for which $g^*_{rl}/z_{no} = \Omega(\log n/\log\log n)$. Since $z\leq z_{no}$ trivially holds, our result implies that Lempel-Ziv compression with overlaps is always at least as good as grammar-compression with run-length rules, and strictly better in some cases.
Formally, we prove the following theorem.

\begin{theorem}
There exists an infinite family of strings for which the ratio between the size of the smallest RLSLP and the length of the LZ77 parse is
$$\frac{g^*_{rl}}{z_{no}} = \Omega\left(\frac{\log n}{\log\log n}\right).$$
\end{theorem}

\section{Prelimaries}
\paragraph{Charikar et al.}
Charikar et al.~\cite{charikar2005smallest} showed a separation between the smallest grammar and the size of the LZ77 parse of a string. 

\begin{lemma}[Charikar et al.]\label{lem:grammar}
There exists an infinite family of strings for which the ratio between size of the smallest grammar and the length of the LZ77 parse is
$$\frac{g^*}{z_{no}} = \Omega\left(\frac{\log n}{\log\log n}\right).$$
\end{lemma}

The proof is based on the following lemma (implicit in the paper) that they proved using a link between grammars and addition chains.
\begin{lemma}[Charikar et al.]\label{lem:lb}
Let $k_1,\dots,k_p$ be a set of distinct positive integers, and consider strings of the form $s = x^{k_1} |_1 x^{k_2} |_2 \ldots |_{p-1} x^{k_p}$, where $k_1$ is the largest of the $k_i$.
Let $p=\Theta(\log k_1)$. There exists an infinite  class of  sequences of integers $k_1,\ldots,k_p$ such that the smallest grammar for $s$ has size 
$$ \Omega\left(\frac{\log^2 k_1}{\log\log k_1}\right).$$
\end{lemma}
Since the LZ77 parse for the string has size $O(p+\log k_1)= O(\log k_1)$ Lemma~\ref{lem:grammar} follows.

\paragraph{Thue-Morse Sequence} The Thue-Morse sequence can be generated 
by starting with $01$ and keep appending the inverse binary negation of the sequence already generated: $$01 \rightarrow  0110 \rightarrow  01101001 \rightarrow   0110100110010110 \rightarrow \ldots $$
The Thue-Morse sequence is overlapfree~\cite{Thue06, Thue12, AS99, Berstel09}, and therefore also cubefree on two symbols~\cite{MH44}. We can obtain a squarefree sequence on three symbols by taking the first difference of the Thue-Morse sequence: 
take the Thue-Morse sequence  $$01101001100101101001011001101001... $$ and form a new sequence in which each term is the difference of two consecutive terms in the Thue-Morse sequence $$\textrm{1 0 -1 1 -1 0 1 0 -1 0 1 -1 1 0 -1 1 -1 0 1 -1 1 0 -1 0 1 0 -1 1 -1 0 1...}$$
This sequence is squarefree (see e.g.~\cite{AS99,Berstel09,Berstel07}). We call this sequence the \difftm sequence.


\section{Separation}

\paragraph{Size of smallest RLSLP} 
Let $t(n)$ be the prefix of length $n$ of the infinite  \difftm sequence. 
Let $k_1,\dots,k_p$ be a set of distinct positive integers, and consider strings of the form $$\hat{s} = t(k_1) |_1 t(k_2) |_2\ldots |_{p-1} t(k_p),$$ where $k_1$ is the largest of the $k_i$.

Since the sequences $t(k_i)$ are squarefree, there is no difference in the size of the smallest grammar and the smallest RLSLP for the string $\hat{s}$.

Let $s = x^{k_1} |_1 x^{k_2} |_2 \ldots |_{p-1} x^{k_p}$. Assume you have a grammar of size $g$ for $\hat{s}$. Replacing all the terminals $(-1, 0, 1)$ by $x$ gives you a grammar for $s$ of  size $g$.
Thus the smallest grammar for $\hat{s}$ must be at least the size of the smallest grammar for $s$.  From Lemma~\ref{lem:lb} we know that there exists exists integers $k_1,\ldots,k_q$ such that the smallest grammar for $s$ has size 
$ \Omega\left(\frac{\log^2 k_1}{\log\log k_1}\right).$ 
It follows that the smallest RLSLP for $\hat{s}$ has size at least $$ \Omega\left(\frac{\log^2 k_1}{\log\log k_1}\right).$$ 

\paragraph{Size of LZ77 parse} 
The LZ77 parse for the Thue-Morse sequence of length $n$ has size $O(\log n)$ \cite{CL06}.
The LZ77 parse  for the \difftm sequence $t(n)$ is at most 2 times larger than the LZ77 parse, $z_t$, for the  corresponding Thue-Morse sequence $t_m(n+1)$: 
We can construct a parse $\hat{z}_t$ of size at most $2|z_t|$ such that each phrase $f$ in $z_t$ gives at most 2 phrases in $\hat{z}_t$. Consider a  phrase $f$ in $z_{t}$. Since phrase $f$ exists earlier in $t_{m}(n+1)$, then the sequence of the differences between the terms in $f$ exists previously in $t$ and we construct a corresponding phrase in $\hat{z}$ of length $|f|-1$. The term denoting the difference between the first position in $f$ and the last position in the previous phrase is in its own phrase. The parse $\hat{z}_t$ has size at most $2|z_t|$ and thus the LZ77 parse of $t(n)$ has size at most $2|z_t|$, since the LZ77 parse is optimal. It follows that the LZ parse of $t(n)$ has size $O(\log n)$.

Now consider the string $\hat{s}$. The LZ77 parse of $\hat{s}$ is then $z_1 |_1 (1,k_2) |_2 \ldots |_{p-1} (1,k_p)$. The size of the parse is  $O(\log k_1+p)= O(\log k_1)$.
The ratio between the smallest RLSLP and the length of the LZ77 parse is therefore 
$$  \Omega\left(\frac{\log k_1}{\log\log k_1}\right) =  \Omega\left(\frac{\log n}{\log\log n}\right).$$

\bibliographystyle{plain}
\bibliography{paper}

\begin{thebibliography}{10}

\bibitem{AS99}
J.-P. Allouche and J.~Shallit.
\newblock The ubiquitous prouhet-thue-morse sequence, 1999.

\bibitem{Berstel09}
Jean {Berstel}, Aaron {Lauve}, Christophe {Reutenauer}, and Franco~V.
  {Saliola}.
\newblock {\em {Combinatorics on words. Christoffel words and repetitions in
  words.}}
\newblock Providence, RI: American Mathematical Society (AMS), 2009.

\bibitem{Berstel07}
Jean Berstel and Dominique Perrin.
\newblock The origins of combinatorics on words.
\newblock {\em European Journal of Combinatorics}, 28(3):996 -- 1022, 2007.

\bibitem{bille2015random}
Philip Bille, Gad~M Landau, Rajeev Raman, Kunihiko Sadakane, Srinivasa~Rao
  Satti, and Oren Weimann.
\newblock Random access to grammar-compressed strings and trees.
\newblock {\em SIAM Journal on Computing}, 44(3):513--539, 2015.

\bibitem{charikar2005smallest}
Moses Charikar, Eric Lehman, Ding Liu, Rina Panigrahy, Manoj Prabhakaran, Amit
  Sahai, and Abhi Shelat.
\newblock The smallest grammar problem.
\newblock {\em IEEE Transactions on Information Theory}, 51(7):2554--2576,
  2005.

\bibitem{CL06}
Sorin Constantinescu and Lucian Ilie.
\newblock The lempel-ziv complexity of fixed points of morphisms.
\newblock In {\em {MFCS}}, volume 4162 of {\em Lecture Notes in Computer
  Science}, pages 280--291. Springer, 2006.

\bibitem{gagie2017optimal}
Travis Gagie, Gonzalo Navarro, and Nicola Prezza.
\newblock Optimal-time text indexing in bwt-runs bounded space.
\newblock {\em arXiv preprint arXiv:1705.10382}, 2017.

\bibitem{kreft2013compressing}
Sebastian Kreft and Gonzalo Navarro.
\newblock On compressing and indexing repetitive sequences.
\newblock {\em Theoretical Computer Science}, 483:115--133, 2013.

\bibitem{lempel1976complexity}
Abraham Lempel and Jacob Ziv.
\newblock On the complexity of finite sequences.
\newblock {\em IEEE Transactions on information theory}, 22(1):75--81, 1976.

\bibitem{MH44}
M.Morse and G.~A. Hedlund.
\newblock Unending chess, symbolic dynamics, and a problem in semigroups.
\newblock 11:1--7, 1944.

\bibitem{nishimoto2016fully}
Takaaki Nishimoto, Shunsuke Inenaga, Hideo Bannai, Masayuki Takeda, et~al.
\newblock Fully dynamic data structure for lce queries in compressed space.
\newblock {\em arXiv preprint arXiv:1605.01488}, 2016.

\bibitem{rytter2003application}
Wojciech Rytter.
\newblock Application of lempel--ziv factorization to the approximation of
  grammar-based compression.
\newblock {\em Theoretical Computer Science}, 302(1-3):211--222, 2003.

\bibitem{Thue06}
A.~Thue.
\newblock {\"U}ber unendliche zeichenreihen.
\newblock {\em Norske vid. Selsk. Skr. Mat. Nat. Kl.}, 7:1--22, 1906.

\bibitem{Thue12}
A.~Thue.
\newblock {\"U}ber die gegenseitige lage gleicher teile gewisser zeichenreihen.
\newblock {\em Norske vid. Selsk. Skr. Mat. Nat. Kl.}, (1):1--67, 1912.

\end{thebibliography}

\end{document}